\documentclass[prc,aps,showpacs,floatfix,nofootinbib,twocolumn,letterpaper]{revtex4}
\usepackage{graphicx}
\usepackage[T1]{fontenc}
\usepackage[latin9]{inputenc}
\usepackage{textcomp}
\usepackage{amsmath}
\usepackage{amssymb}

\def\be{\begin{equation}}
\def\ee{\end{equation}}

\def\u{$^{236}$U}

\def\neck{N_{\rm neck}}

\begin{document}

\title{Decay widths at the scission point in nuclear fission}

\author{G.F.~Bertsch}
\affiliation{Institute for Nuclear Theory and Department of Physics,
University of Washington, Seattle, Washington, USA}
\email{bertsch@uw.edu}
\author{L.M.~Robledo}
\affiliation{Departamento de F\'\i sica Te\'orica, Universidad
Aut\'onoma de Madrid, E-28049 Madrid, Spain}
\affiliation{Center for Computational Simulation,
Universidad Polit\'ecnica de Madrid,
Campus de Montegancedo, Boadilla del Monte, 28660-Madrid, Spain
}
\email{luis.robledo@uam.es}

\begin{abstract}

An outstanding problem in the theory of nuclear fission is understanding
the Hamiltonian dynamics at the scission point.  Here we apply the 
Generator Coordinate Method to calculate
decay widths for pre-scission configurations into the two-fragment continuum.  
Transitions that are allowed under diabatic dynamics can have widths up to
several MeV.  For non-diabatic decays through the pairing interaction,
typical widths to a specific final state channel are 2-3 orders of 
magnitude smaller.
The nucleus \u~is taken as a representative example in the calculations.
\end{abstract}
 
\maketitle

\section{Introduction}

The final step in nuclear fission is the rupture of the neck
between the two nascent fragments, leaving them to interact
only through long-range potential fields.  The quantum dynamics
of this scission process is quite complex and has resisted
a satisfactory description within many-body Hamiltonian theory -- see \cite{sc16} for a recent review.
Here we attempt to construct a fully quantum mechanical treatment in
the framework of self-consistent mean-field theory following the
formulation of Ref. \cite{be19a}.  The configuration
space is constructed by the generator coordinate 
method\footnote{See Refs. \cite{be03,ro19} for reviews of the
method.} (GCM) and interactions are computed at the
nucleon-nucleon level.
Ideally, the GCM basis would be separated into configurations that
are bound under mean-field dynamics and those that will evolve to
separated post-fission fragments under a mean-field Hamiltonian.
The goal of this paper is to make some first estimates of the
transitions between bound states and continuum channels defined
in the same framework by  chains of GCM configurations. 
The GCM constraints on the configurations are their $K$-partitions
and the expectation value of a single-particle operator measuring
the elongation of the system. An important practical question
is
the spacing of the configurations with respect to  elongation
in the chain representing a continuum channel. 
In Ref. \cite{be19a} the quality of the paths was assessed by the overlaps
of the configurations along the chain, but  Hamiltonian dynamics
of their interactions was left for the present exploratory study.  We will also
make use of Ref. \cite{be19b} where the Hamiltonian dynamics of the
GCM configurations leading to separated subsystems was treated
in a general way.  

The transition rate of a bound configuration may be estimated by the 
envelope of its strength function in the
eigenstates of the Hamiltonian in a basis including both the initial state
and the chain of configurations representing the continuum.  For
weak coupling between the initial state and the continuum states,
the decay width can be estimated by 
the Fermi Golden Rule (FGR),
\be
\Gamma = 2 \pi \left|\langle i | H |f\rangle\right|^2\frac{dn_f}{dE}.
\label{FGR}
\ee
Here $|f\rangle $ is a continuum wave function at the same energy
as the initial state $|i\rangle$ and $dn_f/dE$ is the final state
density of states.  

The separation between bound states and those in a
continuum-connected chain is facilitated by using a basis of
Hartree-Fock wave functions in axially symmetric mean fields. 
This permits GCM constraints on the occupation numbers in the
wave function, the $K$-partition mentioned above,  as well as the
familiar shape constraints.  The continuum-connected chain of
GCM configurations is constructed in a diabatic approximation,
namely conserving the $K$-partition along the chain.  However,
the separation between the two kinds of wave function is not
perfect, due in part to the non-orthogonality of GCM basis
states.  Also, as seen in Ref. \cite{be19a}, scission might
require a considerable reorganization of the many-body
wave function even within a given channel. In that case, one
might want to treat the pre-scission side of the chain as the
decaying state.  In such cases where the initial state is not
orthogonal to the continuum channel configurations it is 
necessary to orthogonalize the wave functions before applying
the FGR.  The method used here is via the Lanczos-basis 
strength function \cite{ca05} as described in the Appendix.

In the next section, we present the methodology to
construct continuum final state wave functions in the GCM.
In Sect. III we apply the method to a continuum channel in the
fission of $^{236}$U, partially following the formulation in Ref. 
\cite{be19b}.  Sect. IV presents three examples of calculated 
decay widths to that channel.  We find that the
decays can vary over many orders of magnitude, depending in
part whether the configurations are diabatically connected or
not.
In the last section we discuss possible improvements of the
methodology and the application to physical observables.

\section{Methodology}
The first task in applying the FGR to decay widths is to build 
multi-configuration 
wave functions representing
the decay channels.  The general approach and some numerical considerations
are discussed in Ref. \cite{be19b}.  The configurations are defined as the
Hartree-Fock mean-field states obtained by the GCM
based on  axially symmetric mean fields.   The axial symmetry permits good
$K$ quantum numbers, thus giving the partition by $K$ 
a set of quantum numbers for the many-particle configurations.
In addition, the configurations are labeled by a set of density
constraints as part of the GCM representation.
In previous work we used the mass quadrupole operator $Q_2 =
z^2 - (x^2+y^2)/2$ to generate a  coordinate for the fission path.  Here
we will use instead the relative distance between the two nascent fragments
\be
\label{zrel}
z_{rel} = (z-z_0)\Theta(z-z_0)/A_R + (z_0-z)\Theta(z_0-z)/A_L
\ee 
where $A_R,A_L$ are the number of particles on each side and $z_0$
is the longitudinal position of the dividing plane between the two
nascent fragments.
This field has the advantage that it is exactly the coordinate
needed for the continuum two-fragment wave function of the final state.
Another benefit is that the nuclear part of the force along that coordinate can
be calculated from the properties of the wave function on the dividing
plane \cite{gfb}.

The disadvantage of using Eq. (\ref{zrel}) is that it requires
two parameters, the dividing plane at $z = z_0$ between the two
nascent fragments and the masses on each side of the dividing
plane $A_L$ and $A_R$.  We determine these from the density
distribution immediately before scission.  In particular, $z_0$
is taken as the point where the density on the $z$-axis
$\rho(x=0,y=0,z)$ is minimum. Besides $z_{rel}$, we shall also
employ the neck-size operator  \cite{wa02}
\be
\label{neck}
\hat N_{\rm neck} = \sum_{i=1}^A e^{-(z_i-z_0)^2/a^2}
\ee
to distinguish pre- and post-scission configurations.
The wave functions
are computed using the code HFBaxial \cite{hfbaxial}, which
produces constrained wave functions in a 
Hartree-Fock-Bogoliubov framework.  Here we make use of a technical device to
force the wave functions toward a Hartree-Fock limit.
Namely, an additional constraint is applied to the
fluctuation in proton and neutron particle numbers.  Taking
the constraint as
\be
\langle (\hat N -\langle \hat N \rangle)^2 \rangle = 0.1
\ee
produces HFB wave functions 
that are close to HF configurations. The Gogny D1S energy functional 
is used in this work; other details are the same as in Ref.
\cite{be19a}.

The actual construction of the
multi-configuration eigenstates in the space of GCM configurations is 
quite straightforward, given the matrix elements
of the overlap matrix $S$ and a Hamiltonian matrix $H$ between
configurations 
\be
S_{ij} = \langle i | j \rangle
\ee
\be
H_{ij} = \langle i| {\cal E}| j \rangle,
\ee  
where $|i\rangle$ and $|j\rangle$ are GCM configurations in the space.  
The operator ${\cal E}$ is the energy density functional, treated here
as a Hamiltonian. The standard prescription to handle density dependent
functionals \cite{ro19} is used.  The code GCMaxial \cite{GCM}
is used to compute the $S$ and $H$ matrices from the wave functions generated
by HFBaxial.  The Hamiltonian dynamics is
governed by the equation
\be
i S \frac{d}{dt} \Psi(t) = H \Psi(t);
\ee
the eigenvalue equation is the same with $i d/dt$ replaced by the
eigenvalue.  

There are two crucial assumptions in our procedure 
for constructing continuum channels in the GCM framework.  The first
is 
that the two-fragment final state wave function
factorizes into products of center-of-mass and internal wave functions,
and the second is that the center-of-mass wave functions are Gaussian. Then
the overlaps of final state configurations can be expressed
\be
\langle z_1 | z_2 \rangle = \exp(-(z_1-z_2)^2/4s^2)
\label{olp}
\ee
where $s$ is the size parameter in the Gaussian relative-coordinate
wave function $\psi$:
\be
\label{psi_n}
\psi_n(z_{rel}) = \frac{1}{s^{1/2}\pi^{1/4}}  \exp(- (z_{rel} -z_n)^2/4 s^2) 
\ee
In Ref. \cite{be19b} we analyzed the accuracy of the generated continuum
wave functions for simple model Hamiltonians. The set of GCM configurations
of given $K$-partition form a chain with respect to $\langle z_{rel}\rangle$.
A useful measure in constructing the chain is the overlap distance $\zeta$ between configurations
on the chain.  For a
chain segment containing $N$ configuration the overlap distance between the
two end links is defined
\be
\label{zeta-e}
\zeta_{1,N} = \sum_{n=1}^{N-1} (-\log|\langle n| n+1\rangle|)^{1/2}.
\ee
where $\langle n| n+1\rangle$ are the overlaps of adjacent
configurations.  This definition has the advantage that it is
insensitive to the number of intermediate links and their spacing.
This property is rigorously true for Gaussian overlaps.

An important consideration is how closely to space members of the chain. 
We found that reasonably accurate representations of the GCM Hamiltonian could
be constructed with nearest-neighbor overlaps in the range
$\langle n| n+1\rangle\approx 0.3-0.7$, corresponding to 
overlap distances  $\zeta_{n,n+1} \approx 0.6-1.1$.    

One caveat is that the GCM representation as constrained
here has only a limited capability to approximate finite-momentum
states.  The controlling
parameter is the kinetic energy and its spread associated with the
Gaussian wave packet in the relative coordinate.
Without explicit momentum constraints, the excitation energies
that can be treated are of order 
\be
E \sim  \frac{\hbar^2}{2 M_{red}s^2} 
\ee
or less.  Here $M_{red}$ is the reduced mass associated with the
relative coordinate.
The problem is serious for
fission because the Coulomb field in the final state causes
a large variation in the energies of the configurations as a
function of the separation $r_{rel}$.

In this work we deal with the problem by modifying the $S$ and
$H$ matrix elements associated with the separated fragments  to 
simulate a flat-bottomed potential in $z_{rel}$.
This is implemented as follows.  Partial $S$ and $H$ matrices representing
pre-scission and the closest post-scission configurations are
computed in the usual way. Those matrices are embedding as the
first blocks in larger matrices $S''$ and $H''$ with the additional
entries representing the more distant post-scission configurations.
The additional elements in $S''$ are determined iteratively as
\be
\label{aug-S}
S''_{i,j} = \exp ( -(\zeta_{i,j-1} + \zeta_{j-1,j})^2)
\ee
where $j$ is an added state, $\zeta_{j-1,j}$ is its assumed
distance from the previous state, and $i \le j-1$.  The treatment of the Hamiltonian
matrix element is more complicated.  The diagonal matrix elements
are taken as $H''_{j,j} = H_{k,k}$ where $k$ is the last configuration
included in the full GCM.  The off-diagonal ones need to take into
account the contributions due to non-zero overlaps.
The intrinsic contribution is modeled 
by a quadratic function of $\zeta$, following the Gaussian Overlap
Approximation \cite{be03}.  The resulting parameterization reads
\be
\label{aug-H}
\frac{H''_{i,j}}{S''_{i,j}} = \frac{1}{2}\left(
H_{i,i} + H_{k,k}\right)
+ B\zeta^2_{i,j}.
\ee  
Here $B$ is an introduced parameter.  It is estimated
from the corresponding known elements in $H$ and $S$. 
Alternatively,  $B$ can be deduced from the kinetic Hamiltonian
operator in the final state.  The agreement between the two
ways of estimating it gives a check on the reliability 
of the overall methodology.

There is a technical problem in calculating the Hamiltonian 
matrix element between the initial state and the continuum 
wave function, $\langle i|H| f \rangle$ in the FGR.  
The FGR requires that the 
initial state to be rigorously
orthogonal to the continuum.  This is
certainly the case if the GCM configurations  are HF eigenstates of different
$K$-partitions.  However,
the code to compute $H_{ij}$ makes use of the Balian-Brezin
contraction formula \cite{ba69} which requires the two wave functions
to have a finite overlap\footnote{  
One may use instead the pfaffian based
formula of Ref \cite{be12} which is well defined in this case.}.
Since we actually use the HFB machinery with some residual
pairing, there is no difficulty calculating matrix elements
between configuration connected by pairing.
Thus one can use the code as is for those matrix
elements.  We deal with the non-zero overlaps by explicit orthogonalization
as described in the Appendix.

\section{Glider in the continuum}

\begin{figure}[tb] 
\begin{center} 
\includegraphics[width=1.0\columnwidth]{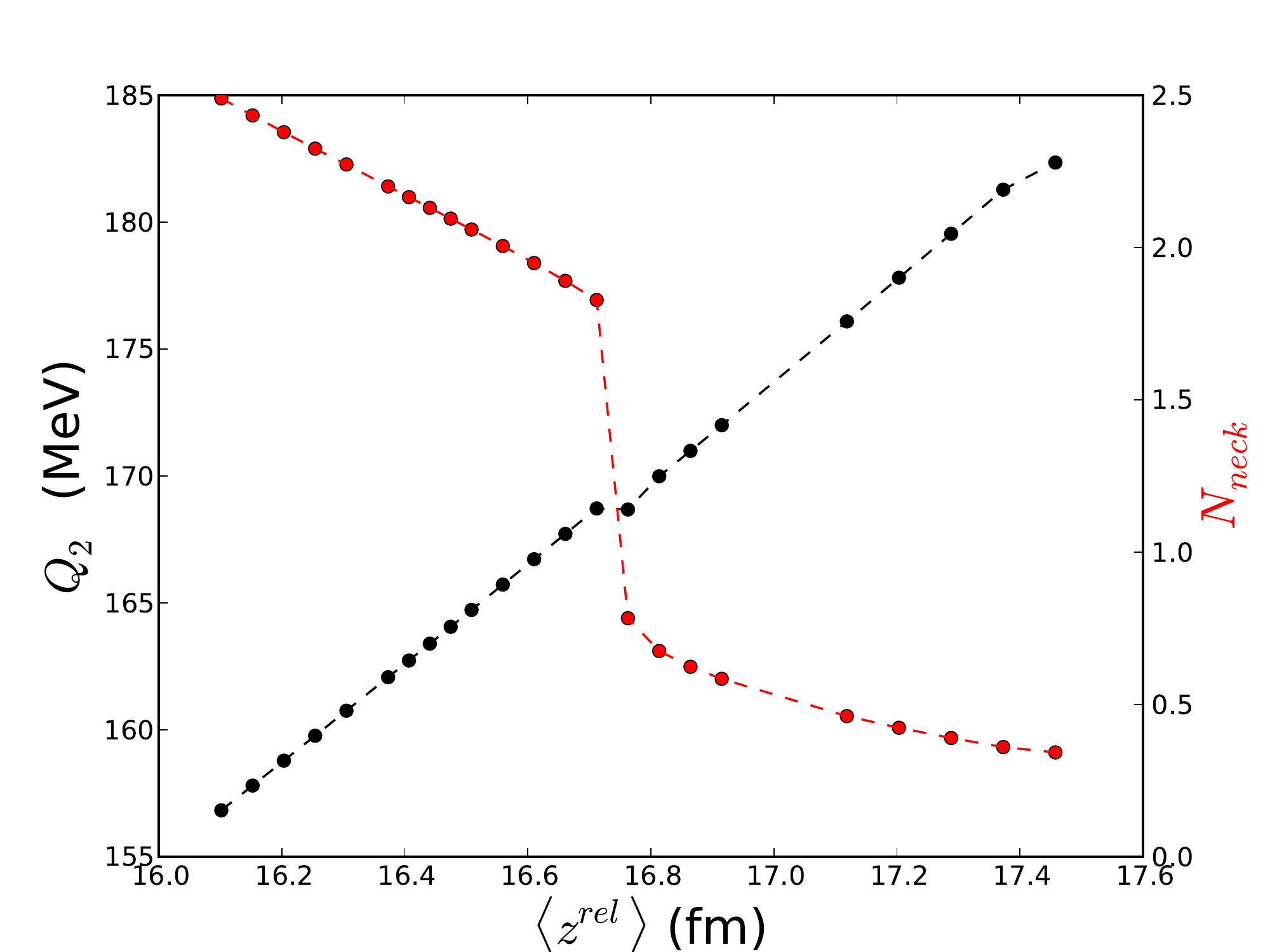} 
\caption{
\label{glider-f}
Black circles:  $Q_2$ of Glider configurations constrained by 
$\langle z \rangle_{rel}$, Eq. (\ref{zrel}).  Red circles:  neck
size $\neck$ (Eq. (\ref{neck})) of the configurations.
}
\end{center}
\end{figure}

Configurations with the $K$-partition called "Glider" in
Refs. \cite{be19a,be18} arise along a GCM-generated
path for the fission of \u. Glider is barely unstable
with respect to fission, so it makes a
good example for the construction of a continuum channel.  
Several characteristics of the Glider scission path are shown in     
Figs.~\ref{glider-f}, 2 and 3.  The configurations in the path
are constrained only by the relative coordinate and the number fluctuation.  
The plot of $\neck$ in Fig. 1 shows that the scission takes
place rather suddenly 
near  $z_{rel} \approx 16.75$ fm.  The figure also shows the mass
quadrupole moment as a function of $z_{rel}$. 
It varies nearly linearly with $z_{rel}$ for the range
of separations shown in the figure, with only a slight offset at
the scission point.  

\begin{figure}[tb] 
\begin{center} 
\includegraphics[width=1.0\columnwidth]{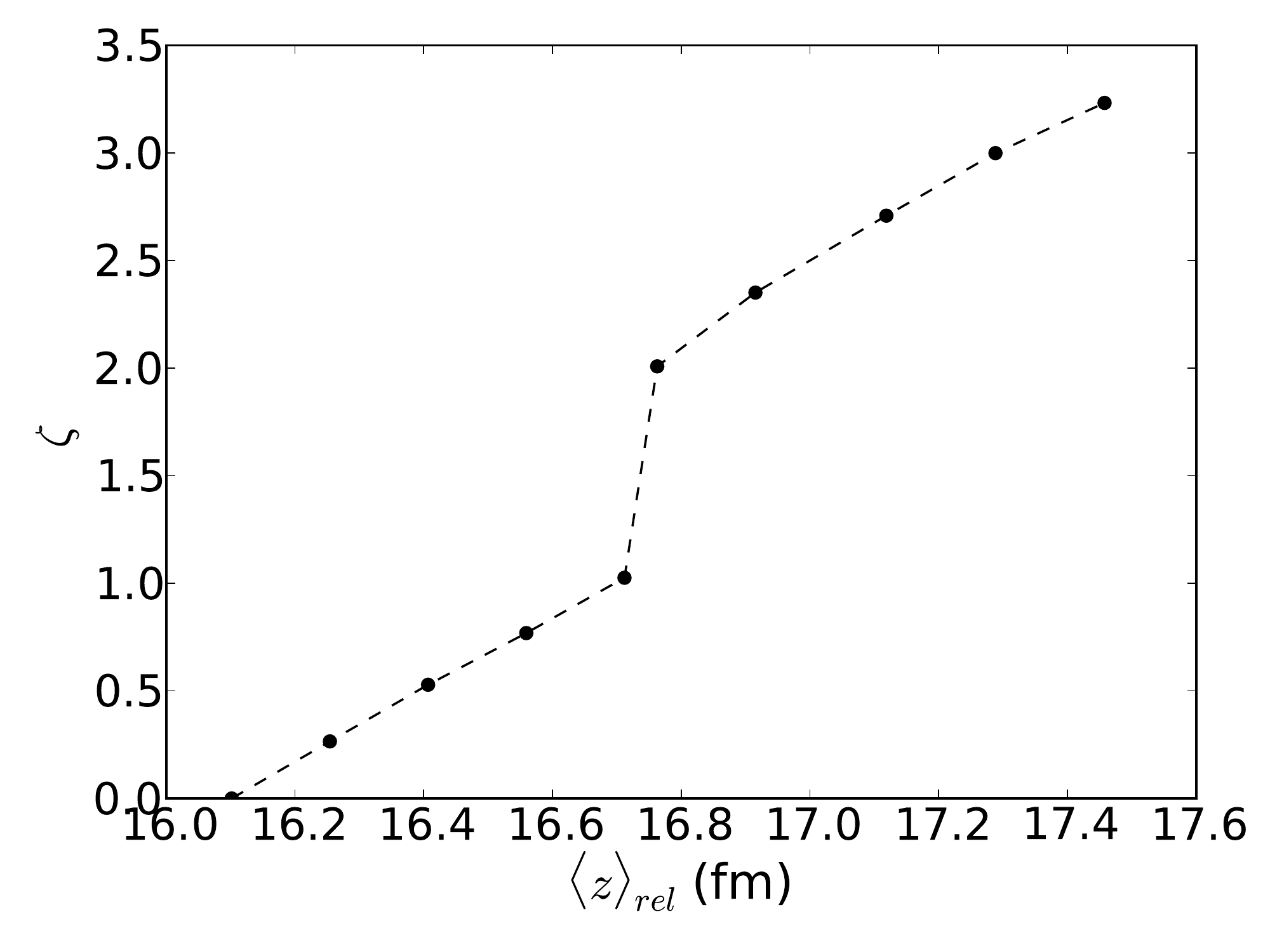} 
\caption{
\label{distance-f}
Overlap distance $\zeta$ along the elongation path 
from $z_{rel} = 16.1$. 
}
\end{center}
\end{figure}

Fig. \ref{distance-f} plots the overlap distance $\zeta$ (Eq.
(\ref{zeta-e})) along the scission path between the coordinates
$z_{rel} = 16.1 $ and 17.4 fm.
One sees that there is a large jump at the scission 
point, showing that the overlaps of the configurations on each side
is much smaller than between neighboring configurations elsewhere along
the chain.  Typical overlaps of the neighboring configurations marked
by circles are in the range 0.85-0.95, while it is only 0.38 across
the scission point.  

\begin{figure}[tb] 
\begin{center} 
\includegraphics[width=1.0\columnwidth]{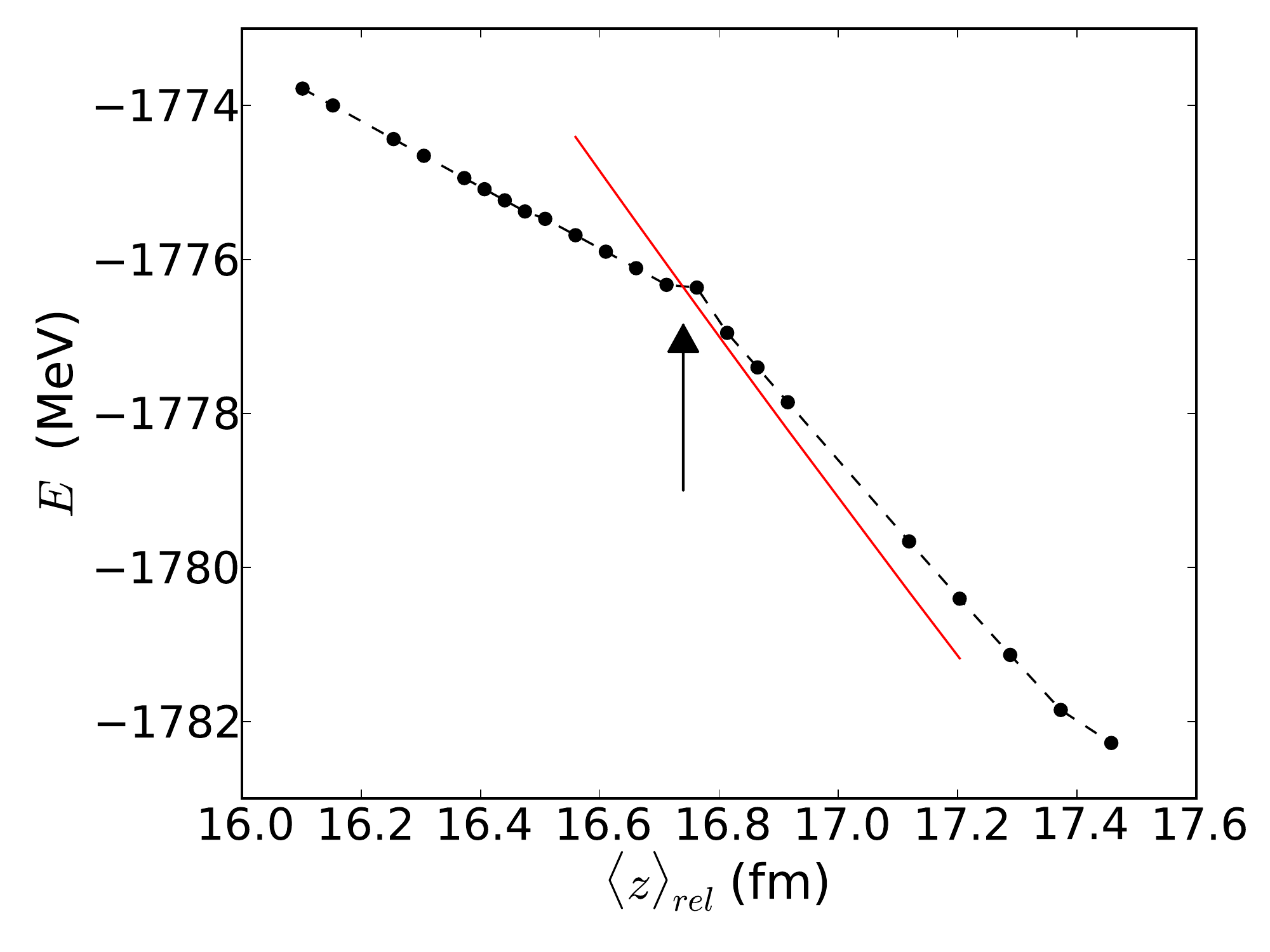} 
\caption{
\label{e-f}
HF energy as a function of the separation between
centers of mass of the two nascent fragments. The
red line shows their Coulomb interaction
approximated as
$V_c = e^2 Z_L Z_R / z_{rel}+ C$ where $C$ is an
offset to cross the energy curve at the scission
point, marked by an arrow.
}
\end{center}
\end{figure}

The HF energies along the scission path are shown in Fig. \ref{e-f}.
The scission point is marked with an arrow.  There one can see
a small offset and change in slope.
Beyond the scission point the slope of the curve should be largely
determined by the Coulomb force between the two fragments.
The red curve shows their Coulomb interaction, offsetted vertically
to facilitate the
comparison with the slope of the energy curve.
The good agreement is promising for the method, 
but could be somewhat misleading in view of the
neglect of the nuclear 
interaction and the shape dependence of the Coulomb field. However, for
well separated fragments the energy nicely follows Coulomb law \cite{wa11}

\begin{figure}[tb] 
\begin{center} 
\includegraphics[width=1.0\columnwidth]{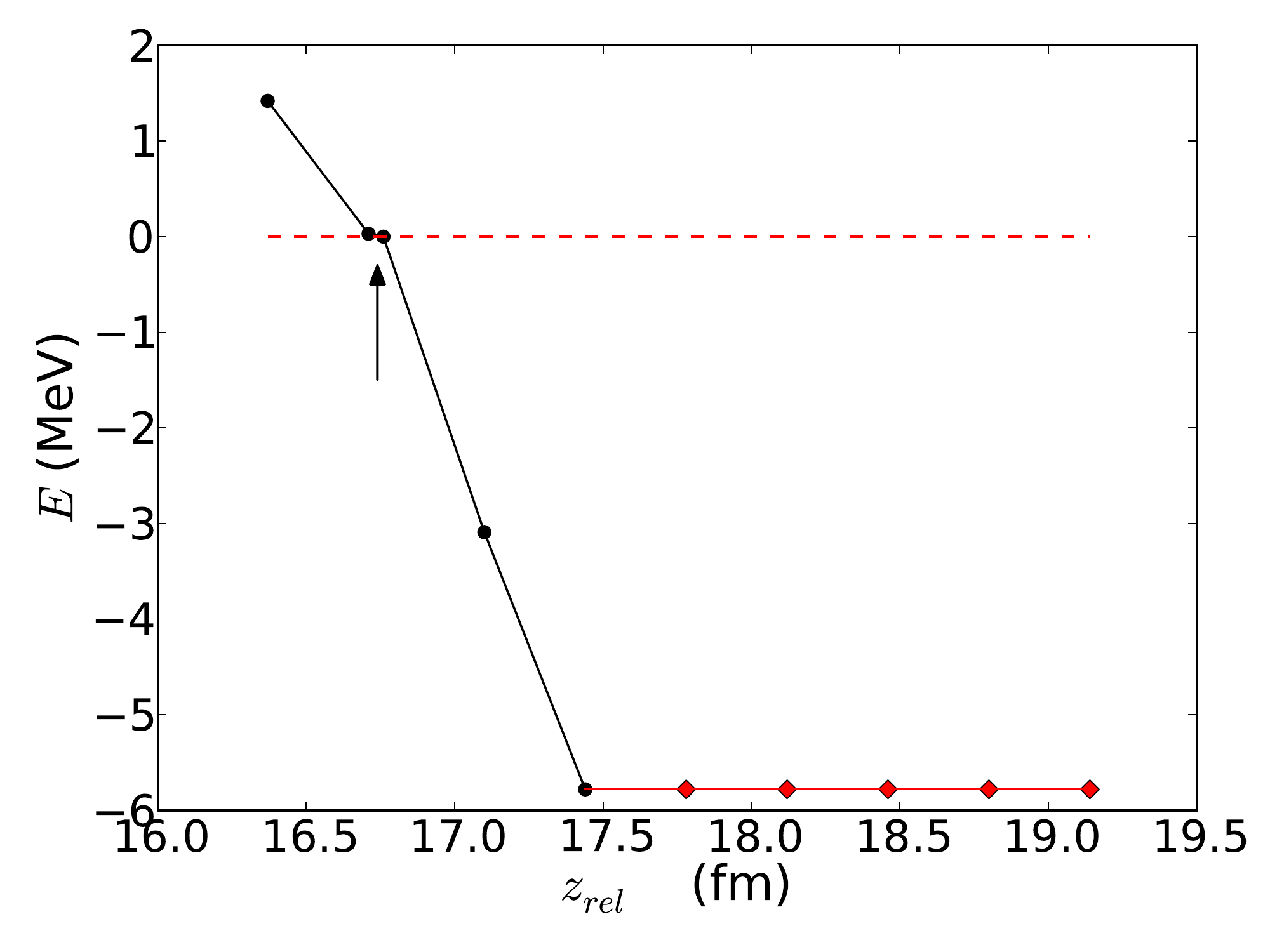} 
\caption{
\label{Vz-f}
Energies of configurations used to build the continuum wave
functions. Black circles are GCM-constrained Glider configurations;
arrow marks the scission point between the second and third state.
Red diamonds are simulated configurations for large separations,
characterized by $S$ and $H$ matrix elements as described in Eq. 
(\ref{aug-S}) and (\ref{aug-H}).  The energy scale is with respect to 
the HF energy of the second configuration.
}
\end{center}
\end{figure}

To construct the truncated continuum wave function we 
start with five glider configurations, composed of two
pre-scission configurations and three just beyond the
scission point.  The farther two are spaced 
at intervals of $\Delta z_{rel} = 0.14$ fm from the
first post-scission configuration. Except for the two on either
side of the scission point, 
the overlaps of neighboring configurations
are about $~0.7$.  This basis is augmented by five
more states at larger separations, with matrix elements 
defined as discussed in the previous section. The positions
and energies of the basis states are shown in Fig. \ref{Vz-f}.
For completeness, the positions, energies, and leading off-diagonal
matrix elements are listed in Table \ref{GCMbasis}.
\begin{table}[htb]  
\begin{center}  
\begin{tabular}{|c|ccccc|}  
\hline  
source & $z_{rel}$ (fm) &   $E$ (MeV) &   $S_{i,i+1}$   & $H_{ij}/S_{ij}$ 
& $a_n$\\
\hline  
GCM & 16.37 & -1774.94 & 0.73& -1776.94 &-0.24\\
    & 16.71 &-1776.32 & 0.38 & -1780.45 &-0.50\\
    & 16.76 &-1776.36 & 0.68 & -1778.45 &-0.21\\
    & 17.10 &-1779.29 & 0.72 & -1782.14 &0.08\\
    & 17.44 &-1782.14 & 0.72 & -1783.57 &0.58\\
\hline
Added    & 17.77 & "& "& "& -0.46\\
   & 18.11 &" &"&" &-0.15\\
    & 18.45 &"& "&" & -0.08\\
    & 18.79 &"&"&" &0.77\\
    & 19.13 &"&-&- &-0.62\\
\hline
\end{tabular} 
\caption{The test states in the space to approximate
the continuum Glider wave function.  The 5 configurations in the
upper part of the table are obtained by the GCM minimization of
the Gogny D1S energy functional.  The lower 5 configurations are
obtained by extrapolating the GCM matrix elements as described in the
text.  The last column shows the amplitudes of the fifth eigenstate
in the spectrum.
\label{GCMbasis}
}
\end{center}  
\end{table}

\begin{figure}[tb] 
\begin{center} 
\includegraphics[width=1.0\columnwidth]{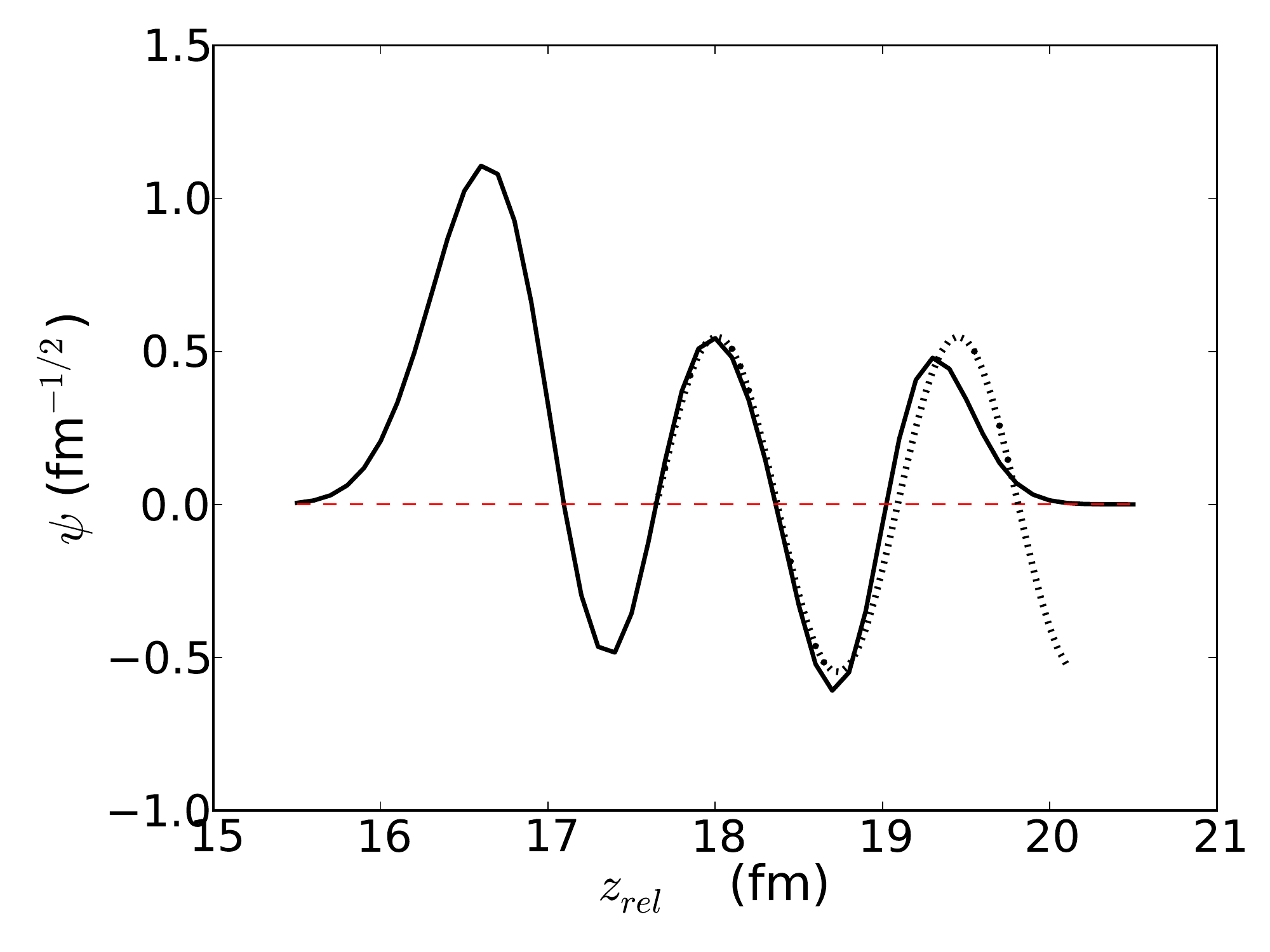} 
\caption{
\label{continuum}  Coordinate space wave function $\psi(z_{rel})$ for the
fifth eigenstate at $E_5=-1776.4$ MeV in the GCM spectrum.  The fit to
$A$ in the asymptotic wave function $A\sin(k z_{rel} + \delta)$ is 
shown as the dotted line in the range $z_{rel} = 17.5 -  18.5$.
}
\end{center}
\end{figure}

After diagonalizing the Hamiltonian, the eigenstate having energy
closest to the initial state is taken as the continuum wave function
of interest.  For our test case here, we take the
fifth state in the 10-dimensional
space, with an eigenenergy of $E_c = -1776.40$ MeV.  This is
close to that of the two configurations near the scission.
The amplitudes $a_{5,n}$ of the eigenfunction are shown in the last column of Table
\ref{GCMbasis}.  The quality of the relative coordinate wave function
can be assessed from its explicit dependence on $z_{rel}$. The
conversion to  $\psi(z_{rel})$ is carried   
assuming that all configurations have the same Gaussian distribution
(Eq. (\ref{psi_n}).  From the overlap of adjacent
configurations (Eq. (\ref{olp})) we determine the parameter $s$ to be
$0.3$ fm.
This is convoluted with the GCM amplitudes given in Table \ref{GCMbasis} to
give the wave function $\psi$ shown in Fig. \ref{continuum}. The wave
function computed this
way comes out properly normalized,
\be
\int |\psi(z_{rel})|^2 d z_{rel} = 1.
\ee

Next we compare with Schr\"odinger wave function for the
relative coordinate, which in this case is a plane wave for
$z_{rel} > 17.44$ fm. 
A fit of the form $\psi(z) = A \sin(k z + \delta)$ is 
shown as the dotted line in 
Fig. \ref{continuum}.  Its parameters are $k=4.36$ fm$^{-1}$ and
$A = 0.55$ fm$^{-1/2}$.  This is to be compared with the energy
to be expected for a plane wave at that momentum in a flat potential.
The energy of the $k=0$ wave function
is lower than the diagonal energy of a GCM configuration due to
the kinetic energy of the GCM wave packet in the relative coordinate.
The energy offset from Eq. (23) of Ref. \cite{be19b} is 
\be
E_0 = \frac{\hbar^2}{4M_{red}s^2}
\ee
which evaluates to 2.05 MeV with the parameters of our system:
$s=0.30$ fm;  $\hbar^2/M_{red} = \hbar^2(A_L + A_R)/m A_L A_R
$; $A_R=136$ and $A_L = 100$.  The kinetic energy
evaluated from the GCM diagonalization is $T_5 = E_5-E_B+E_0
= 7.9 $ MeV.  This is somewhat larger than the kinetic energy for free
particles at the same momentum, $T_{free} = \hbar^2 k^2/2M_{red} = 6.8$ MeV.
This can be interpreted as an (unphysical) effective mass
$M^*/M = 6.8/7.9 = 0.87$. In the
early literature, reproduction of the inertial  masses by the
GCM was considered a challenging problem \cite{pe62}.  If better
accuracy is needed, the GCM wave functions should be constrained
by momentum as well as position.
In any case, the good fit 
of the sinusoidal wave function and the fair reproduction of the 
inertial mass confirms of the adequacy of 
our GCM representation for making rough estimates of decay widths.

\section{Decay width examples}

\begin{figure}[tb] 
\begin{center} 
\includegraphics[width=1.0\columnwidth]{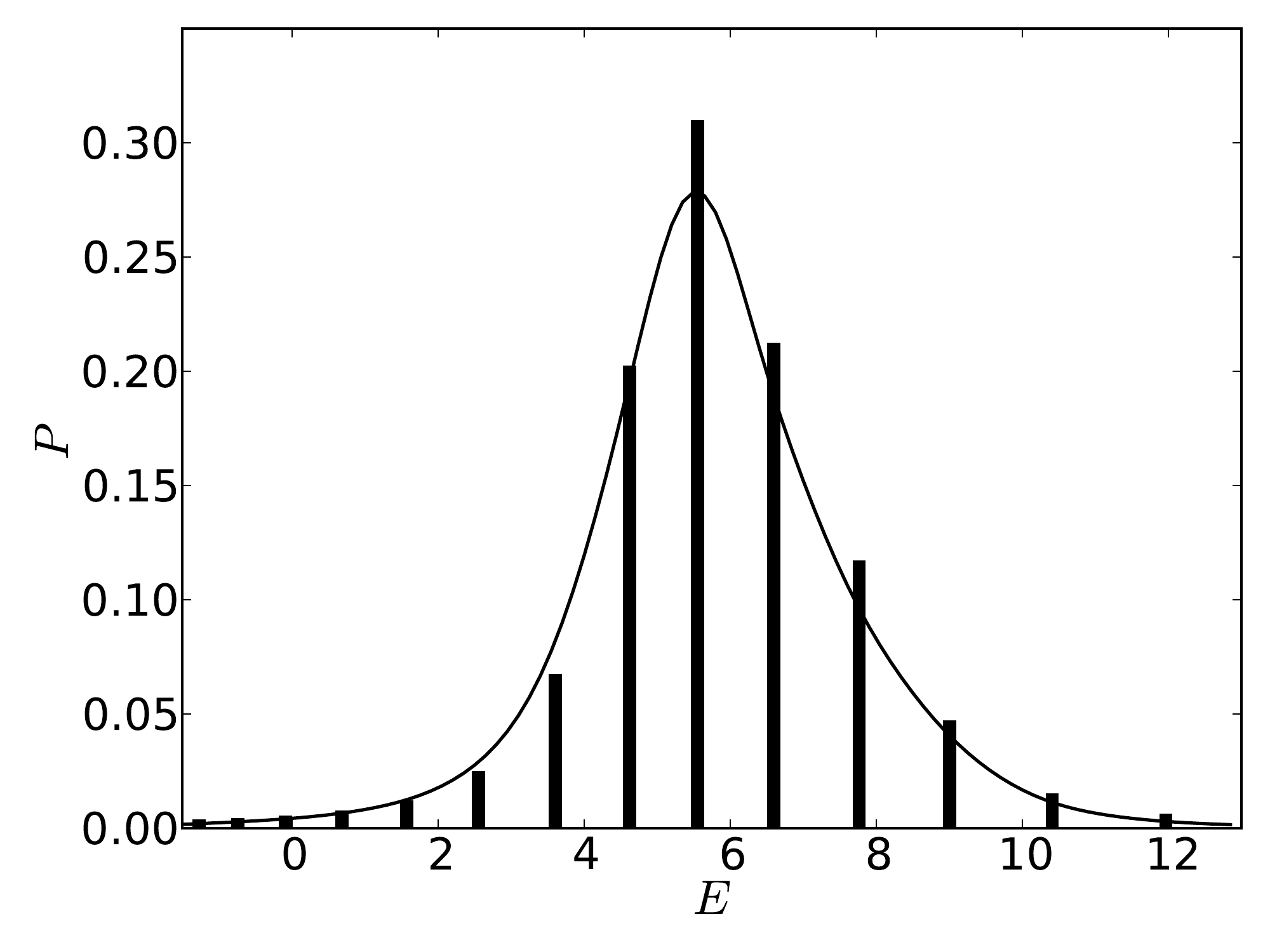} 
\caption{
\label{strength}
The strength function for the Glider 
configuration at $z_{rel} = 16.71$ fm dissolving into the
discretized continuum eigenstates.  The horizontal axis is
the energy of the eigenstates $E_\alpha$ in MeV with respect to 
the flat potential in the external region.
See text for construction
of the discretized basis.
The curve shows the strength function smoothed by convoluting the
discretized strength function with 
Gaussian. 
}
\end{center}
\end{figure}

\begin{figure}[tb] 
\begin{center} 
\includegraphics[width=1.0\columnwidth]{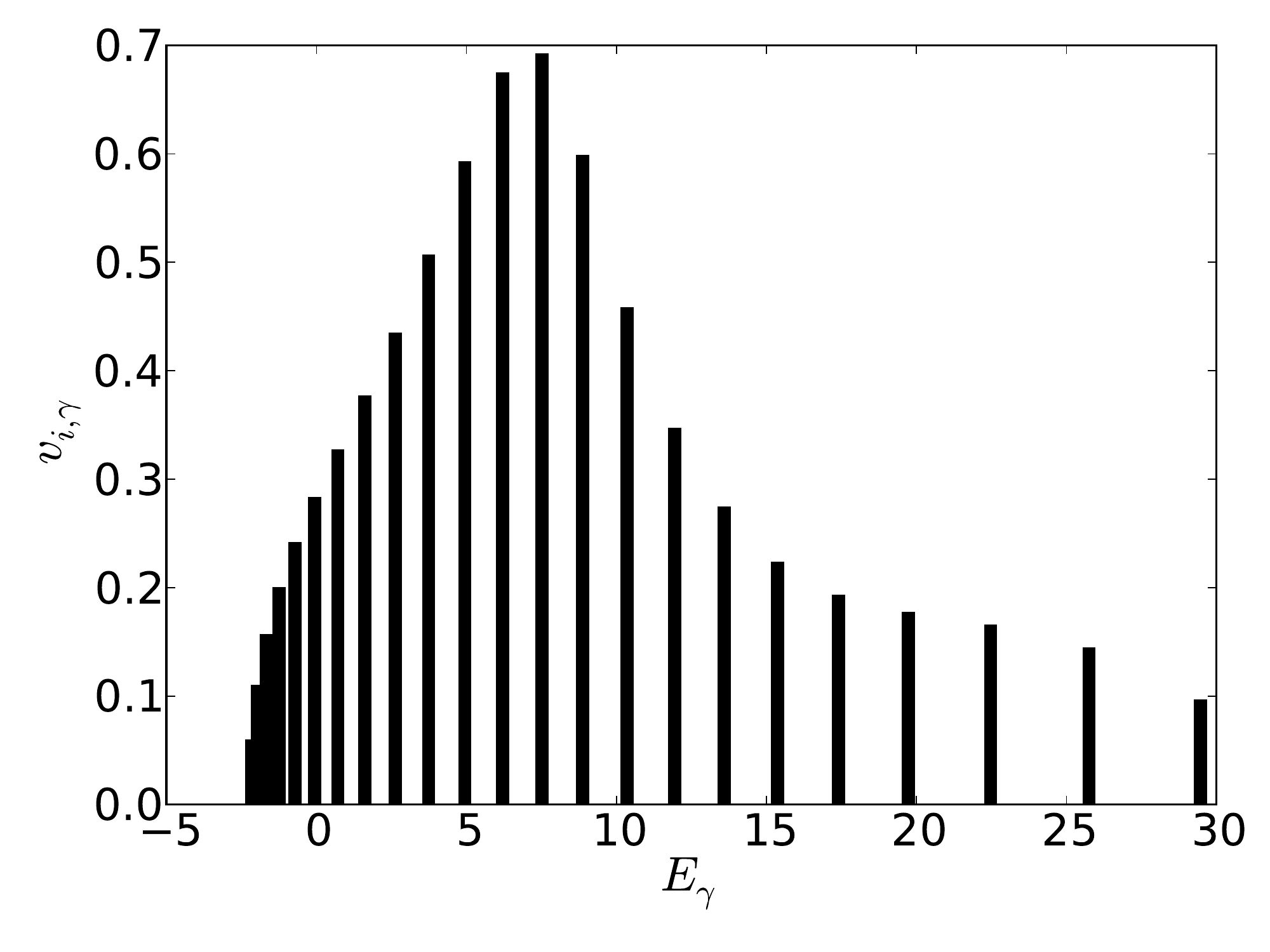} 
\caption{
\label{M2f}
Off-diagonal matrix elements between the initial
state and final states in the $H^\gamma$ matrix.
The horizontal axis is
the energy of the eigenstates $E_\gamma$  with respect to
the flat potential in the external region.
Units are MeV for both axes.
}
\end{center}
\end{figure}

We start with example of a decay through  diabatic dynamics, namely
the decay of the Glider configuration at $z_{rel} = 16.71 $ fm 
to the Glider continuum.  For this case, the decay width is
large enough to assess the mixing directly from the eigenstates
in a fairly small space.  The strength function $P_i$ for a 
configuration in the spectrum 
of eigenfunctions is simply its probability as a function of 
the eigenenergies or a state label for the eigenfunctions.
In the GCM basis this is given by
\be
P_i(\alpha) = \sum_{n}  (S_{i,n} a_{\alpha,n})^2
\label{strength-e}
\ee 
where $i$ is the configuration of interest and $\alpha$ labels
the eigenstates.  This formula satisfies the expected normalization
$\sum_\alpha P_i(\alpha) = 1$.  In the orthonormal basis the 
probability is given by
\be
P_i(\alpha) = \langle \tilde i | \tilde \alpha \rangle^2
\label{strength-e2}
\ee
As mentioned, we take the initial state to be the
Glider configuration at 
$z_{rel} = 16.71$.  The space is augmented with three other 
GCM configurations at $z_{rel} = 16.76, 17.10,$ and 17.44 fm as well
20 configurations constructed with Eq. (14,15) like the ones
in the lower lines of Table \ref{GCMbasis}.  
The resulting strength function is shown
by the vertical bars in Figure \ref{strength}.  There is a 
considerable spread of the strength among the three eigenstates
near $E = -1776$ MeV.
To assign a decay width we first smooth $P_i$ by convoluting
the discrete strengths $P_i(\alpha)$ with a Gaussian spreading
functions.  
Then a decay width can be defined as the full width of the smoothed
strength function at half maximum, $\Gamma_{FWHM}$.  The
extracted value from the Figure is
\be
\Gamma_{\rm FWHM} \approx 3 \,\,\,{\rm MeV}.
\ee
As a check on the modified FGR, we also determined the width
by that method.  Fig. \ref{M2f} shows the off-diagonal
Hamiltonian matrix elements in the first row of  Eq. (\ref{Hg}).
The product is $v^2 = 0.45$ MeV$^2$ at $E_\beta \approx E_i$;
the spacing of energies there is $\Delta E \approx 1.3$ MeV.
Inserting these into Eq. (\ref{fgre}) the estimated decay
with is 
\be
\Gamma_{FGR} = 2.2  \,\,\, {\rm MeV} 
\ee
The agreement is only fair, but one must remember that the
widths are too large to be considered perturbative, and also
the procedure to determine the width from the discrete-basis
strength function was somewhat ad hoc.

Next we treat two cases where the decaying configuration is in
a different $K$-partition from Glider.  In Ref. \cite{be18} we
found that Glider was populated from the bound configuration
Buenavista by two pair jumps.  The orbitals involved are
shown in Table \ref{path-configs}.  The intermediate
configuration labeled ``A" and ``B'' will be treated
as the initial states for the FGR width calculations.  
The coordinate $z_{rel}$ of A and B is set to
16.71 fm, the same as the coordinate of the decaying configuration
in the previous paragraph.  Besides A or B, the space in the
calculation includes all of the GCM states in Table I together
with 20 added configurations in the continuum.  The results
are shown as the last column in Table II.  
\begin{table}[htb]  
\begin{center}  
\begin{tabular}{|c|c|c|c|}  
\hline  
Initial Configuration & Pair jump  &  Final configuration& $\Gamma_{FGR}$\\
\hline
Buenavista &  $(1/2)_p^2 \rightarrow (3/2)_p^2$ &   A &\\
Buenavista & $(1/2)_n^2 \rightarrow (9/2)_n^2$  &   B &\\
A   &  $(1/2)_n^2 \rightarrow (9/2)_n^2$ &  Glider & 5 keV\\
B   &  $(1/2)_p^2 \rightarrow (3/2)_p^2$ & Glider & 60 keV\\
\hline
\end{tabular} 
\caption{
\label{path-configs}
Transitions from Buenavista to Glider via intermediate configurations
A and B.
}
\end{center}  
\end{table}   
One should be cautious in making any quantitative interpretation
of these widths, due to the numerous approximations made to
obtain them.  But we believe that two conclusions can be
drawn already from the three examples.  The first is that
the widths from diabatic dynamics are larger than those
from pairing Hamiltonian by two orders of magnitude or
more.   The
other conclusion is that we should expect large fluctuations in
the widths associated with pairing interaction.
  
\section{Discussion}

We have demonstrated how the GCM framework can be applied to
a fully quantum calculation of the final step in nuclear fission,
namely the rupture of the neck joining the two nascent fragments.
There do not seem to be major obstacles to pursuing
this approach to the point where one can estimate average or
total  decay
widths of the very elongated pre-scission configurations.  We
presented here a calculated decay rate for a configuration
that undergo scission by diabatic dynamics and for two others
configuration that required a residual interaction to reach
the same decay channel.  We hesitate to draw general conclusions
from just these three examples, especially in view of the large
fluctuations in residual interaction matrix elements exhibited
in Table II and also large variations in the overlaps of the
configurations contributing to diabatic decay paths \cite{be19a}.

It is expected that the diabatic dynamics would dominate when
permitted.  For example, the collective masses calculated by
cranking or in other approximations have much larger contributions
from the pairing than the mean-field interaction.
But finding 2-3 orders
of magnitude difference in their contribution to decay widths is 
surprisingly large.  One effect that
could boost the pairing-assisted decays is coherence of the
pairing field in the HFB condensate.  However, that effect is diminished
when the initial wave function contains quasiparticle excitations
that suppress the pairing condensate.  We still don't have
a clear picture of how much thermal excitation energy is present
at the scission point, but the observed presence of odd-even staggering 
in the fragment charge distributions indicates that some pairing
correlations remain.

Experimentally, the finding \cite{be18c} that there are no
systematic fluctuations in the fission cross section on a 1 keV
energy scale indicates that the average total decay rates mediated by
the residual interaction should be considerably larger. All three
of the estimated partial widths were indeed much larger, so the
theory is at least consistent with the observations on this point.

It is also interesting to compare with rates found in the
time-dependent Hartree-Fock-Bogoliubov calculations of 
Ref. \cite{bu15}. In that work it was seen that the early shape 
changes to very
elongated shapes evolved steadily to the scission point,
but the nucleus stops there
for a length of time of the order of 10000 fm/c
before scission occurs. Converting that time to a
decay width gives $\Gamma \approx \hbar/\tau \approx 20 $ keV,
which is between the calculated
decay rates of the A and B configurations into the Glider
channel.  Thus, our microscopic calculation offers some
confirmation of the TDHFB dynamics.

A long-term goal of fission decay width theory is
to calculate not only totals widths
total widths but branching ratios as well.
If the GCM-based theory could be developed to a point
where a representative sample of final continuum channels can be
constructed, it would be possible to estimate branch ratios into the
different channels and thus fluctuations in all observables. 
A good example is the odd-even staggering in mass distributions.
The overall mass yield curves very likely depend mainly on 
statistical dynamics up to populating pre-scission
configurations,  but the final division including pair
breaking requires understanding the scission dynamics.  Another
example is the total kinetic energy distribution, which is 
determined by the access to different exit channels. It might be 
the case that
multiple exit channels compete in the decay of a pre-scission
configuration.  In our first study \cite{be18}, we found that
the pre-scission
configuration Buenavista could connect with two $K$-partitions,
Glider and Bobsled, which exit at quite different kinetic
energies.  

However, there are many problems to be overcome before the theory can be
easily applied to representative samples of configurations.  One
shortcoming of the present formulation is the lack of collective flow in the
GCM parameter space.  The motion of the fragments in the final state is
present in the model space within certain limits.
However, to treat wave functions at fragment separations more than a fermi or so the
kinetic energy would have to be included.  In the GCM, this
could be achieved by a placing a constraint on the momentum
operator.  Operators including currents are also important to assess the 
role of collective flow in pre-scission configurations.

Another problem is a technical one.  Namely, the procedure we
followed to calculate the residual interaction between configurations of $K$-partitions
is specific to the pairing interaction, relying as it does on the non-zero 
overlaps of the HFB wave functions.  The non-pairing residual interaction is
responsible for pair-breaking and increasing (or decreasing) the internal
excitation energy by creating or annihilating quasiparticles.  
It is certainly achievable to 
treat residual interactions of more general form, but this
requires new coding, employing the general algorithm of Ref \cite{be12}
instead
of the more traditional one of Ref. \cite{ba69}.

Another important shortcoming of the method as carried out here is absence of
quasiparticle excitations in the GCM wave functions.  Very likely a
significant fraction of the excitation energy in the primordial fission fragments is in
the form of particle-hole excitations above the base GCM 
configuration \cite{be11}.
It is crucial to know the internal excitation energy of the nascent
fragment to model the subsequent decays emitting neutrons and gamma rays. 

Finally we comment on other fully quantum approaches to fission dynamics.
The time-dependent HFB approximation has been shown to be computationally
feasible \cite{bu15} and interesting results have been obtained from it:
the dynamics is overdamped in the elongation phase and there is a 
long delay at the scission point.  But as a mean-field approximation, the HFB can only give average
behavior and not fluctuations.  Another approach closer to ours is
that of Ref. \cite{go05}.  They consider a large number of GCM
configurations in the HFB approximation and derive a Schr\"odinger
equation for collective GCM variables.  This approach was found to
gives reasonable fluctuations in the mass yields.  But it may not
be so well suited for other quantities, such as the role of
quasiparticle excitations \cite{be11}.  And the particular dynamics
at the scission point seems to us to be beyond the reach of
approaches based on collective shape variables.   
		
\section{Acknowledgment}

We thank W. Younes and U. Mosel for discussions and again to 
W. Younes for contributions to 
computational aspects of the present approach.   The work of LMR was
partly supported by Spanish MINECO grant Nos. FPA2015-65929 and
FIS2015-63770.

\section{Appendix:  Strength function and decay width in the GCM basis}

The strength function $P_i(E)$ of a configuration $i$ mixing with other configurations
is fundamental to the derivation of decay widths.  $P$ is simply given by
the probability of the state $i$ in  the eigenstates $\alpha$ of the Hamiltonian
of the full configuration space,
\be
P_i(E) = \sum_\alpha \langle \tilde i | \tilde \alpha\rangle^2 \delta ( E -E_\alpha).
\label{P_i}
\ee 
where the tilde indicate states defined in the orthonormal
basis\footnote{The tranformation matrix between the GCM and the
orthonormal basis is $S^{1/2}$.}.
Alternately, the Hamiltonian can be
diagonalized directly in the GCM basis  
as the generalized eigenvalue equation
\be
S^{-1} H |\alpha\rangle = E_\alpha |\alpha\rangle.
\ee
Here the $|\alpha\rangle$ are normalized by
\be
\langle \alpha | S | \alpha' \rangle = \sum_{n,n'} a_{n,\alpha}^* S_{n,n'}
a_{n,\alpha'} = \delta_{\alpha,\alpha'}.
\ee
Then the strength function is computed as
\be
P_i(E) = \sum_\alpha \langle i |S |\alpha\rangle^2 \delta ( E -E_\alpha);
\label{P_i-gcm}
\ee
this formulation also satisfies the sum rule   $\int  P_i(E)\,dE = 1$.

When the decays into the continuum is weak, the strength
function approaches a Breit-Wigner shape  $P \sim 1/\left((E-E_i)^2 +
(\Gamma/2)^2\right)^{2}$ 
corresponding to an exponential decay $e^{-\Gamma t}$ in the time domain
Here the FGR can be applied to determine $\Gamma$.  Since Eq. (\ref{FGR})
assumes orthonormality of initial and final states, the safest way to evaluate
the FGR is in an orthonormal basis.  This requires several transformations
from the original GCM representation. The
first step is to convert the vectors to the tilde representation; the
resulting Hamiltonian matrix $H^{\alpha}_{\alpha,\alpha'}$ is Hermitean.
The next transformation
is to tri-diagonalize $H^{\alpha}$ by the Lanczos method using
the state $|\tilde i\rangle$ as the pivot. This yields the tridiagonal
matrix $H^{\beta}$ with basis vectors
${|\tilde i\rangle,|\beta_1\rangle,|\beta_2\rangle,...}$.  
All the states are now orthogonal, but to 
apply the FGR we still need to diagonalize the Hamiltonian in the $\beta$
subspace.  That is carried out by the transformation matrix
\be
U = \left[ \begin{array}{cc}
1  & 0 \\  
0  & U'   
\end{array}
\right]
\ee
where  $U'$ is the transformation matrix to diagonalize $H^\beta$ in the $\beta$
subspace.
The final form of the Hamiltonian is
\be
H^{\gamma} = \left[ \begin{array}{cc}
E_i  & v_{i,\gamma} \\
v^T_{i,\gamma'} & E_\gamma \delta_{\gamma,\gamma'}
\end{array}
\right].
\label{Hg}
\ee 
The matrix elements needed to apply the FGR are the off-diagonal ones in the
first row.  In principle the space should be large  
with a high enough density of  final states to calculate an average
$|v|^2 = \overline {|v_{i,\gamma}|^2}$ over some interval.  Also, the
spacing of continuum states should be uniform enough  
to assign an average spacing $\Delta E$.  Then the FGR can be
evaluated as
\be
\Gamma = \frac{2 \pi}{\Delta E}|v|^2.
\label{fgre}
\ee
Since the size of the spaces is rather small in our examples we
have  taken $|v|^2$ from the matrix element to the state
$\gamma$ that is closest in energy to the initial state.  The corresponding
energy spacing was taken to be
$\Delta E =
(E_{\gamma+1}-E_{\gamma-1})/2$.  This is very approximate, but seems
adequate to estimate the orders of magnitude of the decay widths.

\end{document}